\documentclass[amsmath,amssymb,showkeys,superscriptaddress]{revtex4}
\usepackage{graphicx,epstopdf}
\usepackage{subfig}
\usepackage{mathtools}
\def\xslash{x\!\!\!\slash }

\def\vel{\left|}
\def\ver{\right|}
\usepackage{xcolor}
\usepackage[colorlinks=true, urlcolor=blue, linkcolor=blue, citecolor=green]{hyperref}

\begin{document}

\title{Magnetic dipole moment of $Z_b(10610)$ in light-cone QCD }

\author{U.~\"{O}zdem}%
\email[]{uozdem@dogus.edu.tr}
\affiliation{Department of Physics, Dogus University, Acibadem-Kadikoy, 34722 
Istanbul, Turkey}
\author{K.~Azizi}%
\email[]{kazizi@dogus.edu.tr}
\affiliation{Department of Physics, Dogus University, Acibadem-Kadikoy, 34722 
Istanbul, Turkey}
\affiliation{School of Physics, Institute for Research in Fundamental Sciences (IPM),
P.~O.~Box 19395-5531, Tehran, Iran}

\date{\today}
 
\begin{abstract}
The magnetic dipole moment of the exotic $Z_b(10610)$ state
is calculated within the light cone QCD sum rule method 
using the diquark-antidiquark and
molecule interpolating currents. 
The magnetic dipole moment is obtained 
as $\mu_{Z_b}=1.73\pm 0.63~\mu_N$ in diquark-antidiquark picture and
$\mu_{Z_b}=1.59\pm 0.58~\mu_N$ in the molecular case. 
The obtained results in both pictures together with the results of other theoretical studies on the spectroscopic parameters of the $Z_b(10610)$ state may be useful in determination of the nature and quark organization of  this state. 
\end{abstract}

\maketitle

\section{Introduction}
According to QCD and the conventional quark model, 
not only the standard hadrons,
but also exotic states such as meson-baryon molecules, 
tetraquarks, pentaquarks, glueball and hybrids can exist. 
The first theoretical prediction on the existence of the multiquark 
structures was made by Jaffe in 1976~\cite{Jaffe:1976ih}.
Although it was predicted in the 1970's, 
there was not significant experimental 
evidence of their existence until 2003.
The first observation on the exotic states 
was discovery of $X(3872)$ made by Belle Collaboration~\cite{Choi:2003ue}
 in the decay $B^{+} \rightarrow K^{+}  X(3872) J/\psi \pi^+ \pi^{-}$.  Subsequently, it was confirmed by
BABAR~\cite{Aubert:2004ns}, CDF II~\cite{Acosta:2003zx}, D0~\cite{Abazov:2004kp}, LHCb~\cite{Aaij:2011sn} 
and CMS~\cite{Chatrchyan:2013cld} Collaborations.
The discovery of the X(3872) state turned out to be the 
forerunner of a new direction in hadron physics.
So far, more than twenty exotic states have been observed 
experimentally
[for details, see~\cite{Nielsen:2009uh, Swanson:2006st, 
Voloshin:2007dx,Klempt:2007cp, Godfrey:2008nc,Faccini:2012pj,Esposito:2014rxa, Ali:2017jda}].
The failure of these states to fit the standard particles' structures and violation of some conservation laws such as isospin symmetry,
make these states suitable tools for studying the nonperturbative nature of QCD.

In 2011, Belle Collaboration  discovered two charged 
bottomonium-like states
$Z_b (10610)$ and
$Z_b(10650)$ (hereafter
we will denote these states as $Z_b$ and $Z_b^{\prime}$, 
respectively) in the 
processes $\Upsilon(5S) \rightarrow \pi\pi \Upsilon(nS)$, 
and$\Upsilon(5S) \rightarrow \pi\pi h_b (kP )$~\cite{Belle:2011aa}. 
Here, n = 1, 2, 3 and k = 1, 2.
The masses and widths of the two states have been measured as
             \begin{eqnarray}
              &M_{Z_b}=10607.2 \pm 2~MeV, ~~~~\Gamma_{Z_b}= 18.4 \pm 2.4~MeV, \nonumber\\
              &M_{Z_b^{\prime}}=10652.2 \pm 1.5~MeV,~~~~\Gamma_{Z_b^{\prime}}= 11.5 \pm 2.2~ MeV.\nonumber
             \end{eqnarray}        
             
The analysis of the angular distribution shows that the
quantum numbers of both states are
$I^G(J^P) = 1^+ (1^+)$. Both $Z_b$ and $Z_b^{\prime}$ 
belong to the family
of charged
hidden-bottom states.
Since they are the first observed charged bottomoniumlike states 
and also very close to
the thresholds of $B \bar B^*̄(10604.6 MeV)$ and $B^* \bar B^*(10650.2 MeV)$,
$Z_b$ and $Z_b^{\prime}$ states have attracted attention of
many theoretical groups.
The spectroscopic parameters and decays of $Z_b$ and $Z_b^{\prime}$ states have been studied 
with different models and approaches.
Most of these investigations are based on 
diquark-antidiquark~\cite{Cui:2011fj,Ali:2011ug,Agaev:2017lmc,Wang:2013zra} 
and molecular interpretations~\cite{Cui:2011fj,Bondar:2011ev,Voloshin:2011qa,Zhang:2011jja,
Yang:2011rp,Sun:2011uh,Chen:2011zv,Chen:2011pv,Cleven:2011gp,Cleven:2013sq,
Mehen:2013mva,Wang:2013daa,Wang:2014gwa,Dong:2012hc,Chen:2015ata,Kang:2016ezb,Goerke:2017svb,Dias:2014pva,
Li:2012wf,Li:2012as,Xiao:2017uve,Huo:2015uka}, 
using the analogy to the charm sector. Although the spectroscopic features of these states
have been studied sufficiently, the inner structure of these states have not exactly enlightened.
Different kinds of analyses, such as interaction with the photon can shed light on
 the internal structure of these multiquark states.

A comprehensive analysis of the electromagnetic properties of hadrons ensures 
crucial information on the nonperturbative nature of QCD and their geometric shapes.
The electromagnetic multipole moments contain the spatial distributions of the 
charge and magnetization
in the particle and therefore, 
these observables are directly related to the spatial
distributions of quarks and gluons in hadrons.
In this study, the magnetic dipole moment of the 
exotic state $Z_b$ is extracted by 
using the diquark-antidiquark and
molecule interpolating currents in the framework of the light cone QCD sum rule (LCSR). 
This method has already been successfully applied to study the dynamical and 
statical properties of hadrons for decades such as, form factors, 
coupling constants and multipole
moments. 
In the LCSR, the properties of the particles are characterized in terms of the light-cone distribution amplitudes (DAs)
and the vacuum condensates~[for details, see for instance~\cite{Chernyak:1990ag, Braun:1988qv, Balitsky:1989ry}].

The rest of the paper is organized as follows: In Sec. II, 
the light-cone QCD sum rule for the electromagnetic form factors of
$Z_b$ is applied and its magnetic dipole moment is derived.
Section III, encompasses our numerical analysis and discussion.
The explicit expressions of the photon DAs are moved to the  Appendix A.
%

\section{Formalism}

To obtain the magnetic dipole moment of the $Z_b$ state by using the
LCSR approach, we begin with the subsequent correlation function,
\begin{equation}
 \label{edmn00}
\Pi _{\mu \nu \alpha }(p,q)=i^2\int d^{4}x\,\int d^{4}y\,e^{ip\cdot x+iq \cdot y}\,
\langle 0|\mathcal{T}\{J_{\mu}^{Z_b}(x) J_{\alpha}(y)
J_{\nu }^{Z_b\dagger }(0)\}|0\rangle. 
\end{equation}%

 Here, $J_{\mu(\nu)}$ is the interpolating current of the $Z_b$ state and 
 the electromagnetic current $J_\alpha$ is given as,
\begin{equation}
 J_\alpha =\sum_{q= u,d,b} e_q \bar q \gamma_\alpha q,
\end{equation}
where $e_q$ is the electric charge of the corresponding quark.

From technical point of view, it is more convenient to rewrite the correlation
function by using the external background electromagnetic (BGEM) field,

\begin{equation}
 \label{edmn01}
\Pi _{\mu \nu }(p,q)=i\int d^{4}x\,e^{ip\cdot x}\langle 0|\mathcal{T}\{J_{\mu}^{Z_b}(x)
J_{\nu }^{Z_b\dagger }(0)\}|0\rangle_{F}, 
\end{equation}%
where  F is the external BGEM field and 
$F_{\alpha\beta}= i (\varepsilon_\alpha q_\beta-\varepsilon_\beta q_\alpha)$ with
$q_\alpha$ and $\varepsilon_\beta$  being the four-momentum and polarization
of the BGEM field.
Since the external BGEM field can be made arbitrarily small,
the correlation function in Eq. (\ref{edmn01}) can be acquired by expanding
 in powers of the BGEM field, 
 \begin{equation}
\Pi _{\mu \nu }(p,q) = \Pi _{\mu \nu }^{(0)}(p,q) + \Pi _{\mu \nu }^{(1)}(p,q)+.... ,
\end{equation}
and keeping only terms $\Pi _{\mu \nu }^{(1)}(p,q)$, which corresponds 
to the single photon emission~\cite{Ioffe:1983ju,Ball:2002ps}
(the technical details about the external BGEM field method can be found in~\cite{Novikov:1983gd}).
The main advantage of using the BGEM field approach
relies on the fact that it separates the soft and hard photon emissions in an explicitly gauge 
invariant way~\cite{Ball:2002ps}.
The $\Pi _{\mu \nu }^{(0)}(p,q)$ is the correlation function in the 
absence of the BGEM field, and gives rise to the mass
sum rules of the hadrons, which is not relevant for our case.

After these general remarks, we can now proceed deriving the LCSR
for the magnetic dipole moment of the $Z_b$ state.
The correlation function given in Eq. (\ref{edmn01}) can be obtained 
in terms of hadronic parameters, known as hadronic representation. 
Additionally it can be calculated in terms of the quark and gluon parameters 
in the deep Euclidean region, known as QCD representation.

We can insert a complete set of intermediate hadronic states with the same quantum numbers
as the interpolating current of the $Z_b$ into the correlation function to obtain the hadronic representation. 
Then, by isolating the ground state contributions, 
we obtain the following expression:

\begin{align}
\label{edmn04}
\Pi_{\mu\nu}^{Had} (p,q) = {\frac{\langle 0 \mid J_\mu^{Z_b} \mid
Z_b(p) \rangle}{p^2 - m_{Z_b}^2}} \langle Z_b(p) \mid Z_b(p+q) \rangle_F
\frac{\langle Z_b(p+q) \mid {J^\dagger}_\nu^{Z_b} \mid 0 \rangle}{(p+q)^2 - m_{Z_b}^2} + \cdots,
\end{align}
where  
dots denote the contributions coming from the higher states and
continuum.

The matrix element appearing in Eq. (\ref{edmn04}) can be
written in terms of three invariant form factors as follows~\cite{Brodsky:1992px}:

\begin{align}
\label{edmn06}
\langle Z_b(p,\varepsilon^\theta) \mid  Z_b (p+q,\varepsilon^{\delta})\rangle_F
 &= - \varepsilon^\tau (\varepsilon^{\theta})^\alpha
(\varepsilon^{\delta})^\beta
\Bigg[ G_1(Q^2)~ (2p+q)_\tau ~g_{\alpha\beta}  +
G_2(Q^2)~ ( g_{\tau\beta}~ q_\alpha -  g_{\tau\alpha}~ q_\beta) \nonumber\\
&- \frac{1}{2 m_{Z_b}^2} G_3(Q^2)~ (2p+q)_\tau ~q_\alpha q_\beta  \Bigg]\,,
\end{align}
where $\varepsilon^\tau$ is the polarization vector of the BGEM field;
and $\varepsilon^\theta$ 
and $\varepsilon^{\delta}$ are the 
polarization vectors of the initial and final $Z_b$ states.

The remaining matrix element, that of the interpolating current 
between the vacuum and particle state,
$\langle 0 \mid J_\mu^{Z_b} \mid Z_b \rangle$, is parametrized as

\begin{align}
\label{edmn05}
\langle 0 \mid J_\mu^{Z_b} \mid Z_b \rangle = \lambda_{Z_b} \varepsilon_\mu^\theta\,,
\end{align}
where $\lambda_{Z_b}$ is residue of the $Z_b$ state.

The form factors $G_1(Q^2)$, $G_2(Q^2)$  and $G_3(Q^2)$ can be defined in terms of the charge
$F_C(Q^2)$, magnetic $F_M(Q^2)$ and quadrupole $F_{\cal D}(Q^2)$ form factors as follows 
\begin{align}
\label{edmn07}
&F_C(Q^2) = G_1(Q^2) + \frac{2}{3} (Q^2/4 m_{Z_b}^2) F_{\cal D}(Q^2)\,,\nonumber \\
&F_M(Q^2) = G_2(Q^2)\,,\nonumber \\
&F_{\cal D}(Q^2) = G_1(Q^2)-G_2(Q^2)+(1+Q^2/4 m_{Z_b}^2) G_3(Q^2)\,,
\end{align}
At $Q^2 = 0 $, the form
factors $F_C(Q^2=0)$, $F_M(Q^2=0)$, and $F_{\cal D}(Q^2=0)$ are related to the
electric charge, magnetic moment $\mu$ and the quadrupole moment 
${\cal D}$ as
\begin{align}
\label{edmn08}
&e F_C(0) = e \,, \nonumber\\
&e F_M(0) = 2 m_{Z_b} \mu \,, \nonumber\\
&e F_{\cal D}(0) = m_{Z_b}^2 {\cal D}\,.
\end{align}

 Inserting the
matrix elements in Eqs. (\ref{edmn06}) and (\ref{edmn05})
into the correlation function in Eq. (\ref{edmn04}) and imposing the condition
$q\!\cdot\!\varepsilon = 0$, we obtain the correlation function in terms of
the hadronic parameters as

\begin{align}
\label{edmn09}
 \Pi_{\mu\nu}^{Had}(p,q) &= \lambda_{Z_b}^2  \frac{\varepsilon^\tau}{ [m_{Z_b}^2 - (p+q)^2][m_{Z_b}^2 - p^2]}
 \Bigg[2 p_\tau F_C(0) \Bigg(g_{\mu\nu} -\frac{p_\mu
q_\nu-p_\nu q_\mu}{ m_{Z_b}^2 } \Bigg) \nonumber \\
&+ F_M (0) \Bigg(q_\mu g_{\nu\tau} - q_\nu g_{\mu\tau} +
\frac{1}{m_{Z_b}^2} p_\tau (p_\mu q_\nu - p_\nu q_\mu ) \Bigg)
- \Bigg(F_C(0) + F_{\cal D}(0)\Bigg) {\frac{p_\tau}{m_{Z_b}^2} } q_\mu
q_\nu \Bigg]\,.
\end{align}

To obtain the expression of the correlation function in terms of the quark and gluon parameters,
the explicit form for the interpolating current of the $Z_b$  state needs to be chosen. In this
study, we consider the $Zb$ state  with the quantum numbers  $J^{PC}=1^{+-}$.
Then in the diquark-antidiquark
model the interpolating current $J_{\mu}^{Z_b}$ is 
defined by the following expression 
in terms of quark fields:
\begin{eqnarray}
J_{\mu }^{Z_{b}(Di)}(x) &=&\frac{i\epsilon \tilde{\epsilon}}{\sqrt{2}}\left\{ %
\left[ u_{a}^{T}(x)C\gamma _{5}b_{b}(x)\right] \left[ \overline{d}%
_{d}(x)\gamma _{\mu }C\overline{b}_{e}^{T}(x)\right] 
-\left[ u_{a}^{T}(x)C\gamma _{\mu }b_{b}(x)\right] \left[ \overline{%
d}_{d}(x)\gamma _{5}C\overline{b}_{e}^{T}(x)\right] \right\},
\label{eq:Curr}\\
\end{eqnarray}%
where $C$ is 
the charge conjugation matrix, $\epsilon =\epsilon _{abc}$, $\tilde{\epsilon}=\epsilon _{dec}$;  
and $a,b,...$ are color indices.

One can also construct the interpolating current 
by considering the $Z_b$ as a molecular form of $B \bar B^*$ and $B^* \bar B$ state,

\begin{eqnarray}
 J_{\mu}^{Zb(Mol)}(x)&=&\frac{1}{\sqrt{2}}\Big\{[ \bar d_a(x) i\gamma_5 b_a(x)][\bar b_b(x) \gamma_\mu u_b(x)]
 +[\bar d_a(x) \gamma_\mu b_a(x)][\bar b_b(x) i\gamma_5 u_b(x)]\Big\}.
\end{eqnarray}

After contracting pairs of the light and heavy quark operators, the correlation function becomes:

\begin{eqnarray}
\label{edmn11}
\Pi _{\mu \nu }^{\mathrm{QCD}}(p,q)&=&-i\frac{\epsilon
\tilde{\epsilon}\epsilon^{\prime }\tilde{\epsilon}^{\prime }}{2}
\int d^{4}xe^{ipx} \langle 0 | \Bigg\{  
\mathrm{Tr}\Big[\gamma _{5}\widetilde{S}_{u}^{aa^{\prime }}(x)\gamma _{5}S_{c}^{bb^{\prime }}(x)\Big]
\mathrm{Tr}\Big[\gamma _{\mu }\widetilde{S}_{c}^{e^{\prime }e}(-x)\gamma _{\nu}S_{d}^{d^{\prime }d}(-x)\Big] \notag \\
&&-\mathrm{Tr}\Big[ \gamma _{\mu }\widetilde{S}_{c}^{e^{\prime}e}(-x)\gamma _{5}S_{d}^{d^{\prime }d}(-x)\Big] 
\mathrm{Tr}\Big[ \gamma_{\nu }\widetilde{S}_{u}^{aa^{\prime }}(x)\gamma _{5}S_{c}^{bb^{\prime }}(x)] \nonumber\\
&&-\mathrm{Tr}\Big[\gamma _{5}\widetilde{S}_{u}^{a^{\prime }a}(x)\gamma _{\mu }S_{c}^{b^{\prime}b}(x)\Big]  
\mathrm{Tr}\Big[ \gamma _{5}\widetilde{S}_{c}^{e^{\prime}e}(-x)\gamma _{\nu }S_{d}^{d^{\prime }d}(-x)\Big] \notag \\
&&+\mathrm{Tr}\Big[\gamma _{\nu }\widetilde{S}_{u}^{aa^{\prime }}(x)\gamma _{\mu }S_{c}^{bb^{\prime }}(x)\Big] 
\mathrm{Tr}\Big[\gamma _{5}\widetilde{S}_{c}^{e^{\prime }e}(-x)\gamma_{5}S_{d}^{d^{\prime }d}(-x)\Big]
 \Bigg\}| 0 \rangle_F,
\end{eqnarray}%
in the diquark-antidiquark picture, and

\begin{eqnarray}
\label{neweq}
\Pi _{\mu \nu }^{\mathrm{QCD}}(p,q)&=&-\frac{i}{2}
\int d^{4}xe^{ipx} \langle 0 | \Bigg\{ 
\mathrm{Tr}\Big[\gamma _{5}{S}_{b}^{aa^{\prime }}(x)\gamma _{5}S_{d}^{a^{\prime }a}(-x)\Big]
\mathrm{Tr}\Big[\gamma _{\mu }{S}_{u}^{bb^{\prime }}(x)\gamma _{\nu}S_{b}^{b^{\prime }b}(-x)\Big] \notag \\
&&+\mathrm{Tr}\Big[ \gamma _{5 }{S}_{b}^{aa^{\prime}}(x)\gamma _{\nu}S_{d}^{a^{\prime }a}(-x)\Big] 
\mathrm{Tr}\Big[ \gamma_{\mu }{S}_{u}^{bb^{\prime }}(x)\gamma _{5}S_{b}^{b^{\prime }b}(-x)] \notag \\
&&+\mathrm{Tr}\Big[\gamma _{\mu}{S}_{b}^{aa^{\prime }}(x)\gamma _{5 }S_{d}^{a^{\prime}a}(-x)\Big]  
\mathrm{Tr}\Big[ \gamma _{5}{S}_{u}^{bb^{\prime}}(x)\gamma _{\nu }S_{b}^{b^{\prime }b}(-x)\Big] \notag \\
&&+\mathrm{Tr}\Big[\gamma _{\mu }{S}_{b}^{aa^{\prime }}(x)\gamma _{\nu }S_{d}^{a^{\prime}a}(-x)\Big] 
\mathrm{Tr}\Big[\gamma _{5}{S}_{u}^{bb^{\prime }}(x)\gamma_{5}S_{b}^{b^{\prime }b}(-x)\Big]
 \Bigg\}| 0 \rangle_F,
\end{eqnarray}
in the molecular picture,
where%
\begin{equation*}
\widetilde{S}_{b(q)}(x)=CS_{b(q)}^{\mathrm{T}}(x)C,
\end{equation*}%
with $S_{q}(x)$ and $S_{b}(x)$ being the light and heavy quark propagators, respectively. 
To calculate the correlation functions in QCD representations, 
the light and heavy quark
propagators are required. Their  explicit expressions 
in the $x$-space are given as

\begin{align}
\label{edmn12}
S_{q}(x)=i \frac{{\xslash}}{2\pi ^{2}x^{4}} 
- \frac{\langle \bar qq \rangle}{12}\Big( 1+  \frac{m_0^2 x^2}{16}\Big)   
-\frac {i g_s }{32 \pi^2 x^2} G^{\mu \nu}(x) \Big[\rlap/{x} 
\sigma_{\mu \nu} 
+  \sigma_{\mu \nu} \rlap/{x}
 \Big],
\end{align}%
and
%
\begin{eqnarray}
\label{edmn13}
&&S_{b}(x)=\frac{m_{b}^{2}}{4 \pi^{2}} \Bigg[ \frac{K_{1}\Big(m_{b}\sqrt{-x^{2}}\Big) }{\sqrt{-x^{2}}}
+i\frac{{\xslash}~K_{2}\Big( m_{b}\sqrt{-x^{2}}\Big)}
{(\sqrt{-x^{2}})^{2}}\Bigg]
-\frac{g_{s}m_{b}}{16\pi ^{2}} \int_0^1 dv\, G^{\mu \nu }(vx)\Bigg[ (\sigma _{\mu \nu }{\xslash}
  +{\xslash}\sigma _{\mu \nu })\frac{K_{1}\Big( m_{b}\sqrt{-x^{2}}\Big) }{\sqrt{-x^{2}}}\nonumber\\
&&+2\sigma_{\mu \nu }K_{0}\Big( m_{b}\sqrt{-x^{2}}\Big)\Bigg],
\end{eqnarray}%
where $K_i$ are the second kind Bessel functions, $v$ is line variable and $G^{\mu\nu}$ is the gluon field strength tensor. 

The correlation function includes  different types of contributions.
In first case, one of the free quark 
propagators in Eqs.~(\ref{edmn11}-\ref{neweq}) is replaced by
\begin{align}
\label{sfree}
S^{free} \rightarrow \int d^4y\, S^{free} (x-y)\,\rlap/{\!A}(y)\, S^{free} (y)\,,
\end{align}
where $S^{free}$ is the first term of the light or heavy quark propagators
and the remaining three propagators are replaced with the full quark propagators.
The LCSR calculations are most conveniently done in the fixed-point gauge. For electromagnetic field,
it is defined by $x_\mu A^\mu =0$. 
In this gauge, the electromagnetic potential is given by
\begin{align}
\label{AAA}
 &A_\alpha = -\frac{1}{2} F_{\alpha\beta}y^\beta 
   = -\frac{1}{2} (\varepsilon_\alpha q_\beta-\varepsilon_\beta q_\alpha)\,y^\beta.
\end{align}
The Eq. (\ref{AAA}) is plugged into Eq. (\ref{sfree}), as a result of which we obtain
 \begin{align}
  S^{free} \rightarrow -\frac{1}{2} (\varepsilon_\alpha q_\beta-\varepsilon_\beta q_\alpha)
  \int\, d^4y \,y^{\beta}\, 
  S^{free} (x-y)\,\gamma_{\alpha}\,S^{free} (y)\,,
 \end{align}

After some calculations for $S_q^{free}$ and $S_b^{free}$  
we get
\begin{eqnarray}
&& S_q^{free}=\frac{e_q}{32 \pi^2 x^2}\Bigg(\varepsilon_\alpha q_\beta-\varepsilon_\beta q_\alpha\Bigg)
 \Bigg(\xslash\sigma_{\alpha \beta}+\sigma_{\alpha\beta}\xslash\Bigg),\nonumber\\
&& S_b^{free}=-i\frac{e_b m_b}{32 \pi^2}
\Bigg(\varepsilon_\alpha q_\beta-\varepsilon_\beta q_\alpha\Bigg)
\Bigg[2\sigma_{\alpha\beta}K_{0}\Big( m_{b}\sqrt{-x^{2}}\Big)
 +\frac{K_{1}\Big( m_{b}\sqrt{-x^{2}}\Big) }{\sqrt{-x^{2}}}
 \Bigg(\xslash\sigma_{\alpha \beta}+\sigma_{\alpha\beta}\xslash\Bigg)\Bigg].
\end{eqnarray}

In second case one of the light quark 
propagators in Eqs.~(\ref{edmn11}-\ref{neweq}) are replaced by
\begin{align}
\label{edmn14}
S_{\alpha\beta}^{ab} \rightarrow -\frac{1}{4} (\bar{q}^a \Gamma_i q^b)(\Gamma_i)_{\alpha\beta},
\end{align}
and the remaining propagators are full quark propagators including 
the perturbative as well as the nonperturbative contributions.
Here as an example, we give a short detail of the calculations of the QCD representations.
 In second case for simplicity, we only consider the first term in Eq.~(\ref{edmn11}),
\begin{align}
\label{QCDES}
 &&\Pi _{\mu \nu }^{\mathrm{QCD}}(p,q)=-i\frac{\epsilon
\tilde{\epsilon}\epsilon^{\prime }\tilde{\epsilon}^{\prime }}{2}
\int d^{4}xe^{ipx} \langle 0 |  
\mathrm{Tr}\Big[\gamma _{5}\widetilde{S}_{u}^{aa^{\prime }}(x)\gamma _{5}S_{b}^{bb^{\prime }}(x)\Big]
\mathrm{Tr}\Big[\gamma _{\mu }\widetilde{S}_{b}^{e^{\prime }e}(-x)\gamma _{\nu}S_{d}^{d^{\prime }d}(-x)\Big]
|0\rangle_F+...
 \end{align}
 
 By replacing one of light propagators with the expressions in Eq. (\ref{edmn14})
 and making use of 
 \begin{align}
  \bar q^a(x)\Gamma_i q^{a'}(0)\rightarrow \frac{1}{3}\delta^{aa'}\bar q(x)\Gamma_i q(0),
 \end{align}
the Eq. (\ref{QCDES}) takes the form

 \begin{align}
\label{QCDES2}
 &\Pi _{\mu \nu }^{\mathrm{QCD}}(p,q)=-i\frac{\epsilon
\tilde{\epsilon}\epsilon^{\prime }\tilde{\epsilon}^{\prime }}{2}
\int d^{4}xe^{ipx} \Bigg\{ 
\mathrm{Tr}\Big[\gamma _{5}\Gamma_i \gamma _{5}S_{b}^{bb^{\prime }}(x)\Big]
\mathrm{Tr}\Big[\gamma _{\mu }\widetilde{S}_{b}^{e^{\prime }e}(-x)\gamma _{\nu}S_{d}^{d^{\prime }d}(-x)\Big]
\frac{1}{12}\delta^{aa'}  \nonumber\\
&+\mathrm{Tr}\Big[\gamma _{5}\widetilde{S}_{u}^{aa^{\prime }}(x)\gamma _{5}S_{b}^{bb^{\prime }}(x)\Big]
\mathrm{Tr}\Big[\gamma _{\mu }\widetilde{S}_{b}^{e^{\prime }e}(-x)\gamma _{\nu}\Gamma_i\Big]
\frac{1}{12}\delta^{dd'}\Bigg\} \langle \gamma(q) |\bar q(x)\Gamma_i q(0)|0\rangle 
+...,
 \end{align}
where $\Gamma_i = I, \gamma_5, \gamma_\mu, i\gamma_5 \gamma_\mu, \sigma_{\mu\nu}/2$.
Similarly, when a light propagator interacts with the photon, 
a gluon may be released from one of the remaining three propagators. 
The expression obtained in this case is as follows:
\begin{align}
\label{QCDES3}
 &\Pi _{\mu \nu }^{\mathrm{QCD}}(p,q)=-i\frac{\epsilon
\tilde{\epsilon}\epsilon^{\prime }\tilde{\epsilon}^{\prime }}{2}
\int d^{4}xe^{ipx} \Bigg\{
\mathrm{Tr}\Big[\gamma _{5}\Gamma_i \gamma _{5}S_{b}^{bb^{\prime }}(x)\Big]
\mathrm{Tr}\Big[\gamma _{\mu }\widetilde{S}_{b}^{e^{\prime }e}(-x)\gamma _{\nu}S_{d}^{d^{\prime }d}(-x)\Big]
\Big[\Big(\delta^{ab}\delta^{a'b'}  -\frac{1}{3}\delta^{aa'}\delta^{bb'}\Big)\nonumber\\
 &+\Big(\delta^{ae}\delta^{a'e'}-\frac{1}{3}\delta^{aa'}\delta^{ee'}\Big)
 +\Big(\delta^{ad}\delta^{a'd'}-\frac{1}{3}\delta^{aa'}\delta^{dd'}\Big)\Big]\nonumber\\
 &+\mathrm{Tr}\Big[\gamma _{5}\widetilde{S}_{u}^{aa^{\prime }}(x)\gamma _{5}S_{b}^{bb^{\prime }}(x)\Big]
\mathrm{Tr}\Big[\gamma _{\mu }\widetilde{S}_{b}^{e^{\prime }e}(-x)\gamma _{\nu}\Gamma_i\Big]
\Big[\Big(\delta^{db}\delta^{d'b'}  -\frac{1}{3}\delta^{dd'}\delta^{bb'}\Big)
+\Big(\delta^{de}\delta^{d'e'}-\frac{1}{3}\delta^{dd'}\delta^{ee'}\Big)
\nonumber\\
&+\Big(\delta^{ad}\delta^{a'd'}-\frac{1}{3}\delta^{aa'}\delta^{dd'}\Big)\Big]
\Bigg\}\frac{1}{32} \langle \gamma(q) |\bar q(x)\Gamma_i G_{\mu\nu}(vx) q(0)|0\rangle 
+...,
 \end{align}
where we inserted
\begin{align}
\label{QCDES5}
 \bar q^a(x)\Gamma_i G_{\mu\nu}^{bb'}(vx) q^{a'}(0)\rightarrow \frac{1}{8}\Big(\delta^{ab}\delta^{a'b'}
 -\frac{1}{3}\delta^{aa'}\delta^{bb'}\Big)\bar q(x)\Gamma_i G_{\mu\nu}(vx) q(0).
\end{align}

As is seen, there appear matrix
elements such as $\langle \gamma(q)\vel \bar{q}(x) \Gamma_i q(0) \ver 0\rangle$
and $\langle \gamma(q)\vel \bar{q}(x) \Gamma_i G_{\mu\nu}(vx)q(0) \ver 0\rangle$,
representing the nonperturbative contributions. 
These matrix elements can be expressed in terms 
of photon DAs and wave functions with definite
twists, whose expressions are given in Appendix A. 
%
The QCD representation of the correlation function is obtained by using 
Eqs.~(\ref {edmn11}-\ref {QCDES5}). 
Then, the Fourier transformation
is applied to transfer expressions in x-space to the momentum space.

The sum rule for the  magnetic dipole moment are obtained by 
matching the expressions 
of the correlation function in terms
of QCD parameters and its expression in terms of the hadronic parameters, 
using their spectral
representation. 
To eliminate the contributions of the excited and continuum states 
in the spectral representation of the correlation function, a double Borel
transformation with respect to the variables $p^2$ and $(p + q)^2$ is applied. 
After the transformation, these contributions are
exponentially suppressed. 
Eventually, we choose the structure  
$(\varepsilon.p)( p_\mu q_\nu- q_\mu p_\nu)$ for the magnetic dipole moment and obtain 
\begin{align}
 &\mu^{Di} =\frac{e^{m_{Z_b}^2/M^2}}{\lambda_{Z_b}^2  m_{Z_b}^2}\Bigg[\Pi_1+\Pi_2\Bigg],\\
 &\mu^{Mol} =\frac{e^{m_{Z_b}^2/M^2}}{\lambda_{Z_b}^2 m_{Z_b}^2}\Bigg[\Pi_3+\Pi_4\Bigg].
\end{align}
 
The explicit forms of the functions that appear in the
above sum rules are given as follows:
\begin{eqnarray}
 \Pi_1&=&\frac{3m_b^4}{256 \pi^6}(e_u-e_d)\Bigg\{32N[3,3,0]-2M^2 N[3,3,1]
  -16m_b N[3,4,1]+m_b M^2 N[3,4,2]\Bigg\}
 \nonumber\\
 &&-\frac{m_b^2 \langle g_s^2 G^2\rangle}{9216 \pi^6}(e_u-e_d)
 \Bigg(-M^2 N[1,1,0]+2m_b N[1,2,0]\Bigg)\nonumber\\
 &&-\frac{m_b^2 \langle g_s^2 G^2\rangle}{147456 \pi^6}\Bigg(2m_b M^2 N[1,2,1]
 +\pi^2 \langle \bar qq \rangle\Big(16 N[1,2,1]+5 N[1,2,2]\Big)\Bigg)\nonumber\\
 &&+\frac{m_b^4 \langle g_s^2 G^2\rangle}{294912 \pi^6}(e_u-e_d)
 \Bigg(16N[1,3,1]-M^2N[1,3,2]\Bigg)\nonumber\\
 &&-\frac{m_b^2 \langle g_s^2 G^2\rangle}{294912 \pi^6}(e_u-e_d)\Bigg(128 N[2,2,0]
 -8(2m_b^2+M^2)N[2,2,1]+m_b^2 M^2N[2,2,2]\Bigg)\nonumber\\
  &&-\frac{m_b^3}{98304 \pi^6}(e_u-e_d)\Bigg(
  16\Big(\langle g_s^2G^2 \rangle -192 \pi^2 m_b \langle \bar qq \rangle\Big)N[2,3,1]
+M^2\Big(13\langle g_s^2G^2 \rangle -960 \pi^2 m_b \langle \bar qq \rangle\Big)N[2,3,2]
  \nonumber\\
  &&-\frac{m_b^3 m_0^2 \langle \bar qq \rangle  }{384 M^8 \pi^4}(e_u+e_d)\Bigg(64m_b^6FlP[-3,4,0]
  -48m_b^4FlP[-2,4,0]+12m_b^2FlP[-1,4,0]-FlP[0,4,0]\Bigg)\nonumber\\
  &&-\frac{m_b m_0^2 \langle g_s^2G^2 \rangle \langle \bar qq \rangle }{36864 M^8 \pi^4}
  (e_u-e_d)\Bigg(16m_b^4FlP[-1,2,0]-8m_b^2FlP[0,2,0]+FlP[1,2,0]\Bigg),\\
  \nonumber\\
  \nonumber\\
   && \Pi_2=-\frac{m_b \langle g_s^2G^2 \rangle \langle \bar qq \rangle^2 }{2592 M^{10} \pi^2}
 (e_u-e_d)(m_0^2-M^2)I_3[h_\gamma]\Bigg(4m_b^2FlNP[0,1,0]-FlNP[-1,1,0]\Bigg)\nonumber\\
 &&+\frac{m_b m_0^2 \langle g_s^2G^2 \rangle \langle \bar qq \rangle^2}{10368 M^{10}\pi^2}
 (e_u-e_d)I_3[h_\gamma]\Bigg(4m_b^2FlNP[2,1,1]-FlNP[3,1,1]\Bigg)\nonumber\\
 &&-\frac{f_{3\gamma} m_0^2 \langle g_s^2G^2 \rangle \langle \bar qq \rangle}{110592 M^{10}\pi^2}\Big[
 -(4e_u-3e_d)\psi^a(u_0)+2e_d I_3[\psi^\nu]\Big]\Bigg(16m_b^4FlNP[1,2,1]-8m_b^2FlNP[2,2,1]\nonumber\\
 &&+FlNP[3,2,1]\Bigg)\nonumber\\
 &&-\frac{m_b m_0^2 \langle g_s^2G^2 \rangle \langle \bar qq \rangle}{995328 M^{12}\pi^2}
 (8e_u-5e_d)(A(u_0)+8I_3[h_\gamma])\Bigg(16m_b^4FlNP[2,3,2]
 -8m_b^2FlNP[3,3,2]+FlNP[4,3,2]\Bigg)\nonumber\\
 &&+\frac{m_b \langle \bar qq \rangle^2 }{497664 M^{12} \pi^2}(8e_u-5e_d)\Bigg[
 -\langle g_s^2G^2 \rangle\Big(-(5m_0^2-2M^2)A(u_0)
 -2 m_0^2 \chi M^2  \varphi_\gamma(u_0)\Big) 
 -8\Big\{-5m_0^2 \langle g_s^2G^2 \rangle\nonumber\\
&& +2\Big(M^2 \langle g_s^2G^2 \rangle +432 m_b^2 m_0^2 M^2 \Big) I_3[h_\gamma]\Big\}\Bigg]
\Bigg(16m_b^4 FlNP[0,3,1]-8m_b^2FlNP[1,3,1]+FlNP[2,3,1]\Bigg)\nonumber\\
&&+\frac{m_b}{165888 M^{10}\pi^4}\Bigg[(e_u-e_d)f_{3\gamma}\Big(7M^2 \langle g_s^2G^2 \rangle
-576\pi^2 m_b \langle \bar qq \rangle (m_0^2-M^2)\Big)\psi^a(u_0) 
+72\pi^2 \langle \bar qq \rangle^2(m_0^2-M^2)\nonumber\\
&&\Big(-2e_u I_1[\tilde S]+e_d\Big(3I_2[\mathcal{T}_1]-3I_2[\mathcal{T}_2]
-5I_2[\tilde S]\Big)\Big)\Bigg]\Bigg(-64m_b^6FlNP[3,4,0]
+48m_b^4FlNP[2,4,0]\nonumber\\
&&-12m_b^2FlNP[1,4,0]+FlNP[0,4,0]\Bigg)\nonumber\\
&&-\frac{m_b m_0^2  \langle \bar qq \rangle }{9216 M^{10}\pi^4}\Bigg[-8(e_u-e_d)m_b f_{3\gamma}\psi^a(u_0)
- \langle \bar qq \rangle \Big(-2e_u I_1[\tilde S]+e_d\Big(3I_2[\mathcal{T}_1]
-3I_2[\mathcal{T}_2]-5I_2[\tilde S]\Big)\Big)\Bigg]\nonumber\\
&&\Bigg(64m_b^6FlNP[-1,4,1]-48m_b^4FlNP[0,4,1]
+12m_b^2FlNP[1,4,1]-FlNP[2,4,1]\Bigg)\nonumber
  \end{eqnarray}
   \begin{eqnarray}
&&+\frac{\langle \bar qq \rangle}{1990656 M^{12}\pi^4}\Bigg[
\langle g_s^2G^2 \rangle\Bigg\{2e_u\Bigg(-3M^4\Big(16m_b^4FlNP[1,3,0]-8m_b^2FlNP[0,3,0]+FlNP[-1,3,0]\Big)\nonumber\\
&&-32\pi^2 m_b  \langle \bar qq \rangle (5m_0^2-4M^2)\Big(16m_b^4FlNP[2,3,0]-8m_b^2FlNP[1,3,0]+FlNP[0,3,0]\Big)\Bigg)\nonumber\\
&&+e_d\Bigg(40\pi^2 m_b \langle \bar qq \rangle(5m_0^2-4M^2) 
\Big(16m_b^4FlNP[2,3,0]-8m_b^2FlNP[1,3,0]+FlNP[0,3,0]\Big)\nonumber\\
&&+3M^4\Big(-48m_b^6FlNP[2,3,0]+56m_b^4FlNP[1,3,0]-19m_b^2FlNP[0,3,0]+2FlNP[-1,3,0]\Big)\Bigg)\Bigg\}A(u_0)\nonumber\\
&&+8m_b\Bigg\{-4(8e_u-5e_d)\pi^2 M^2 \chi \langle g_s^2G^2 \rangle \langle \bar qq \rangle(m_0^2-M^2)\varphi_\gamma(u_0)
-3(40e_u-43e_d)m_b M^4 \langle g_s^2G^2 \rangle\nonumber\\
&&+8(8e_u-5e_d)\pi^2  \langle g_s^2G^2 \rangle \langle \bar qq \rangle(5m_0^2-M^2)
-432(e_u-e_d)\pi^2 m_b^2 M^2 \langle \bar qq\rangle (m_0^2-4M^2)\Bigg\}I_3[h_\gamma]\Bigg]\nonumber\\
&& \Big(16m_b^4FlNP[2,3,0]-8m_b^2FlNP[1,3,0]+FlNP[0,3,0]\Big),
\end{eqnarray}

\begin{eqnarray}
 \Pi_3&=&\frac{9 m_b^4}{1024 \pi^6}(e_u-e_d)\Bigg\{
 32N[3,3,0]-2M^2N[3,3,1]-16m_bN[3,4,1]+m_b M^2N[3,4,2] \Bigg\}\nonumber\\
 &&-\frac{m_b^3}{32768 \pi^6}(e_u-e_d)(\langle  g_s^2 G^2 \rangle +48 \pi^2 m_b \langle \bar qq\rangle )
 \Bigg( 16N[2,3,1]+5M^2N[2,3,2]  \Bigg)\nonumber\\
 &&-\frac{m_b^3 m_0^2 \langle \bar qq\rangle }{512 M^8 \pi^2}(e_u-e_d)\Bigg(64m_b^64FlP[-3,4,0]
 -48m_b^4FlP[-2,4,0]+12m_b^2FlP[-1,4,0]+FlP[0,4,0]\Bigg),
\end{eqnarray}

and

\begin{eqnarray}
 \Pi_4&=&\frac{m_b^2 \langle g_s^2 G^2\rangle \langle \bar qq\rangle}{294912 \pi^4}\Bigg[
 e_u\Big(-3I_1[\mathcal{S}]-2I_1[\tilde S]\Big)
 +e_d\Big(3I_2[\mathcal{S}]+2I_2[\tilde S]\Big)\Bigg]\Bigg(M^2N[1,2,2]-8N[1,2,1]\Bigg)\nonumber\\
 &&+\frac{3m_b^2 \langle \bar qq\rangle }{64 \pi^4}(e_u-e_d)I_3[h_\gamma]
 \Bigg(M^2N[2,3,2]-8N[2,3,1]\Bigg)
 +\frac{3m_b^4 f_{3\gamma}}{128 \pi^4}(e_u+e_d)\psi^a(u_0)\Bigg(M^2N[3,3,2]-8N[3,3,1]\Bigg)\nonumber\\
 &&-\frac{m_b^3 m_0^2 \langle \bar qq\rangle^2 }{96 M^{10}\pi^2}(e_u-e_d)I_3[h_\gamma]
 \Bigg(64 m_b^6FlNP[0,3,1]-48m_b^4FlNP[1,3,1]+FlNP[2,3,1]\Bigg)\nonumber\\
 &&-\frac{m_b^3 m_0^2 f_{3\gamma}\langle \bar qq\rangle^2 }{1536 M^{10}\pi^2}(e_u-e_d)\psi^a(u_0)
 \Bigg(64m_b^6FlNP[-1,4,1]-48FlNP[0,4,1]+12m_b^2FlNP[1,4,1]\nonumber\\
 &&-FlNP[2,4,1]\Bigg)\nonumber\\
 &&+\frac{m_b f_{3\gamma}}{18432 M^{10}\pi^4}(e_u+e_d)\Bigg[\Big(
 -M^2 \langle g_s^2 G^2\rangle-48\pi^2m_b  \langle \bar qq\rangle(m_0^2-M^2)\Big)\psi^a(u_0)\Bigg]
 \Bigg(-64m_b^6FlNP[3,4,0]\nonumber\\
 &&+48m_b^4FlNP[2,4,0]-12m_b^2FlNP[1,4,0]+FlNP[0,4,0]\Bigg)\nonumber\\
 &&-\frac{m_b \langle \bar qq \rangle}{1152 M^{10}\pi^4}(e_u-e_d)\Bigg[\Big(
 -M^2 \langle g_s^2 G^2\rangle-48\pi^2m_b  \langle \bar qq\rangle(m_0^2-M^2)\Big)I_3[h_\gamma]\Bigg]
 \Bigg(-16m_b^4FlNP[2,3,0]\nonumber\\
 &&-8m_b^2FlNP[1,3,0]+FlNP[0,3,0]\Bigg).
\end{eqnarray}
where, $m_b$ is the mass of the b quark, $e_q$ is the corresponding electric charge,
$\chi$ is the magnetic susceptibility of the quark condensate,
$m_0^2 = \langle \bar q g \sigma_{\alpha\beta} G^{\alpha\beta} q \rangle /\langle \bar qq \rangle $, 
$\langle \bar qq \rangle$ and $\langle g_s^2 G^2\rangle$ are quark and gluon condensates, respectively.
 
The functions $N[n,m,k]$,~$FlP[n,m,k]$,~$FlNP[n,m,k]$,
~$I_1[\mathcal{A}]$, 
~$I_2[\mathcal{A}]$ and $I_3[\mathcal{A}]$ are
defined as:
\begin{align}
 N[n,m,k]&=\int_0^\infty dt\int_0^\infty dt'~ 
 \frac{e^{-m_b/2(t+t')}}{t^n~(\frac{m_b}{t}+\frac{m_b}{t'})^k~ t'^m },\nonumber\\
 FlP[n,m,k]&= \int_{4m_b^2}^{s_0} ds \int_{4m_b^2}^s dl~ \frac{e^{-l^2/\phi}~ 
 l^n~ (l-s)^m}{(4m_b^2-l)^2~\phi^k},\nonumber\\
 FlNP[n,m,k]&=  \int_{4m_b^2}^{s_0} ds \int_{4m_b^2}^s dl~ \frac{e^{-l^2/\beta}~ 
 l^n~ (l-s)^m}{(l-2m_b^2)~\beta^k},\nonumber\\
   I_1[\mathcal{A}]&=\int D_{\alpha_i} \int_0^1 dv~ \mathcal{A}(\alpha_{\bar q},\alpha_q,\alpha_g)
 \delta(\alpha_ q +\bar v \alpha_g-u_0),\nonumber\\
  I_2[\mathcal{A}]&=\int D_{\alpha_i} \int_0^1 dv~ \mathcal{A}(\alpha_{\bar q},\alpha_q,\alpha_g)
 \delta(\alpha_{\bar q}+ v \alpha_g-u_0),\nonumber\\
 I_3[\mathcal{A}]&=\int_0^1 du~ A(u),\nonumber\\
 \end{align}
 where
\begin{align}
 &\beta=4\,l\,M^2-16\,m_b^2M^2,\nonumber\\
 &\phi=8\,l\,M^2-32\,m_b^2M^2.\nonumber
\end{align}

The functions $ \Pi_1 $ and $ \Pi_3 $ indicate the case that one of the 
quark propagators enters the perturbative interaction with the photon and 
the remaining three propagators are taken as full propagators.
The functions $ \Pi_2 $ and $ \Pi_4 $ show the contributions that one of the light quark propagators 
enters the nonperturbative interaction with the photon and 
the remaining three propagators are taken as full propagators.
The reader can  find some details about the calculations 
such as Fourier and Borel transformations as well as continuum 
subtraction in Appendix C of Ref.~\cite{Ozdem:2017jqh}.

As we already mentioned, the calculations have been done in the fixed-point gauge, $x_\mu A^\mu =0$, for simplicity.  In order to show whether
 our results are gauge invariant or not we examine the Lorentz gauge, $\partial_\mu A^\mu =0$. In this gauge, the electromagnetic vector potential is written as 
\begin{align}
 &A_\mu(x) = \varepsilon_\mu e^{-iq.x},
\end{align}
with $ \varepsilon_\mu q^\mu =0 $.
In this gauge, the corresponding gauge invariant electromagnetic field strength tensor is written  as 
 \begin{align}
  F_{\mu \nu}=i(\varepsilon_\mu q_\nu-q_\mu \varepsilon_\nu)e^{-iq.x}.
 \end{align}
We repeat all the calculations in this gauge and find the same results 
for the magnetic dipole moment of the state under consideration. 
Therefore the results obtained in the present study are gauge invariant.

\section{Numerical analysis and Conclusion}

In this section, we numerically analyze the results 
of calculations for magnetic dipole moment of the $Z_b$ state.
We use $m_{Z_b}= 10607.2 \pm  2 MeV$, $\overline{m}_b(m_b) = (4.18_{-0.03}^{+0.04})~GeV$~\cite{Olive:2016xmw}, 
$f_{3\gamma}=-0.0039~GeV^2$~\cite{Ball:2002ps},  
$\langle \bar qq\rangle(1\,GeV) =(-0.24\pm0.01)^3\,GeV^3$ \cite{Ioffe:2005ym},
$m_0^{2} = 0.8 \pm 0.1~GeV^2$, $\langle g_s^2G^2\rangle = 0.88~ GeV^4$~\cite{Nielsen:2009uh} and 
$\chi(1\,GeV)=-2.85 \pm 0.5~GeV^{-2}$~\cite{Rohrwild:2007yt}. 
To evaluate a numerical
prediction for the magnetic moment, 
we need also specify the values of the residue of the $Z_b$ state.
The residue is obtained from the mass sum rule as 
$\lambda_{Z_b}= m_{Z_b} f_{Z_b}$ with 
$f_{Z_b}=(2.79^{+0.55}_{-0.65})\times10^{-2}~GeV^4$~\cite{Agaev:2017lmc} 
for diquark-antidiquark picture and
$\lambda_{Z_b}=0.27\pm 0.07~GeV^5$~\cite{Zhang:2011jja} for molecular picture.
The parameters used in the photon DAs are given 
in Appendix A, as well.

The estimations for the  magnetic dipole moment of the $Z_b$ state depend on two
 auxiliary parameters; the continuum threshold $s_0$ and Borel mass parameter $M^2$.
 The continuum threshold is not completely an arbitrary parameter, and there are some physical restrictions for it.
  The $s_0$ signals the scale at which, the excited states and continuum start to
contribute to the correlation function.
The working interval for this parameter is chosen such that
the maximum pole contribution is acquired and the results relatively
weakly depend on its choices. Our numerical calculations lead 
to the interval $[119-128]~GeV^2$ for this parameter. 
 %
 The Borel parameter can vary in the interval that the results 
 weakly depend on it according to the standard prescriptions.
 The upper bound of it is found demanding the maximum 
 pole contributions and its lower bound is found the convergence 
 of the operator product expansion and exceeding of the perturbative
 part over nonperturbative contributions.
Under these constraints, the working region of the Borel parameter 
is determined as $ 15~GeV^2 \leq M ^ 2 \leq 17~GeV^2 $.

 In Fig. 1, we plot the dependency of the magnetic dipole moment of the $Z_b$ state on $M^2$ 
 at different fixed values of the continuum threshold.
   From the figure we observe that the results considerably depend on the variations of the Borel parameter.
The  magnetic dipole moment is 
stable under variation of $s_0$ in its working region.
In Fig. 2, we show the contributions of $ \Pi_1 $, $ \Pi_2 $, $ \Pi_3 $ and $ \Pi_4 $ functions
 to the results obtained at average value of $s_0$ with respect to the Borel mass parameter.
It is clear that $ \Pi_1 $ is dominant in the results 
obtained when using diquark-antidiquark current but 
$ \Pi_3 $ is dominant while using the molecular current. 
The contribution of the $ \Pi_2 $ and $ \Pi_4 $ functions seems to be almost zero.
 When the results are analyzed in detail, almost (95-97)\% 
 of the total contribution 
 comes from the perturbative part 
 and the remaining (3-5)\% belongs to the nonperturbative contributions.

\begin{figure}
\centering
\subfloat[]{\label{fig:TMagMMsq.eps}\includegraphics[width=0.4\textwidth]{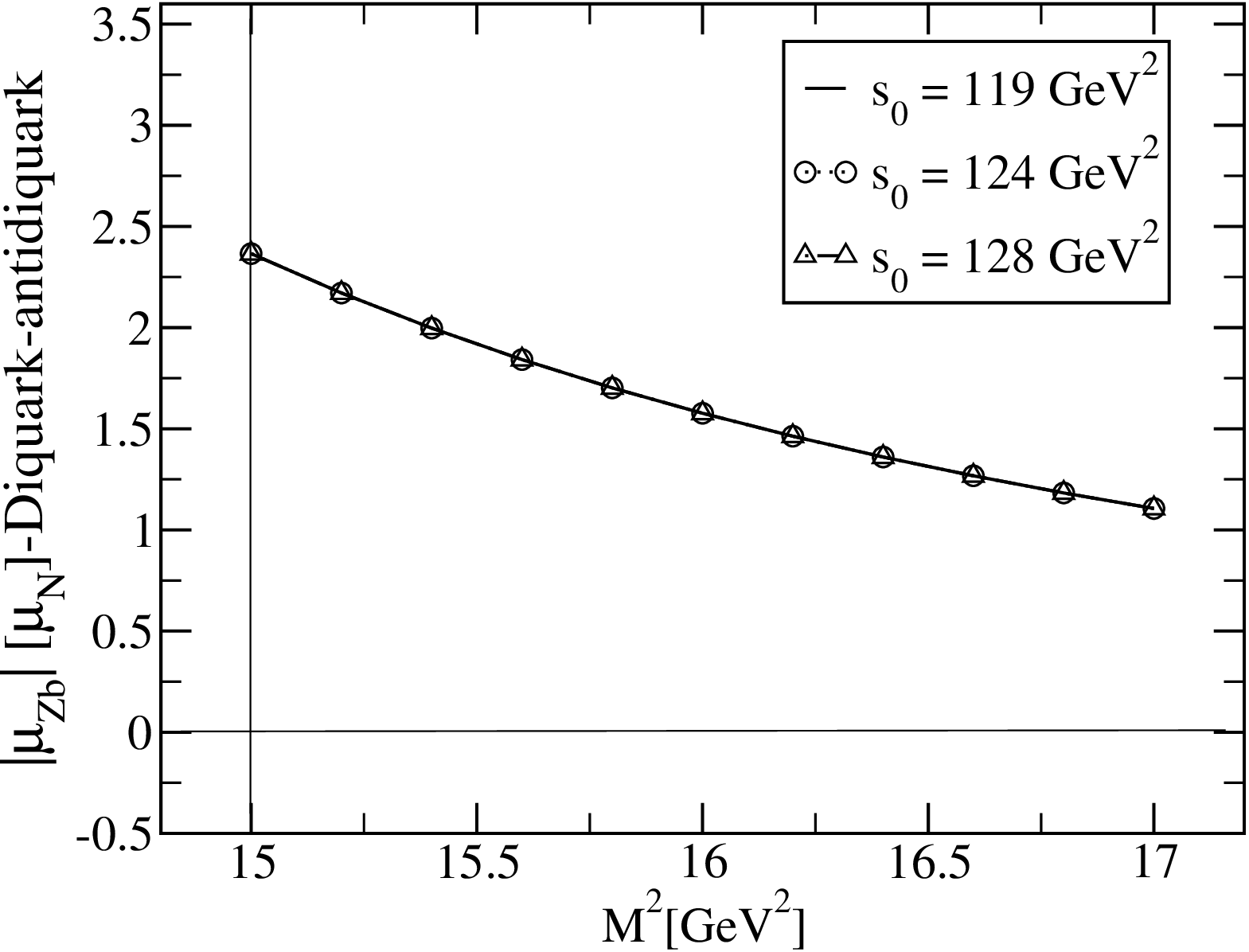}}\\
\vspace{0.5cm}
\subfloat[]{\label{fig:MMagMMsq.eps}\includegraphics[width=0.4\textwidth]{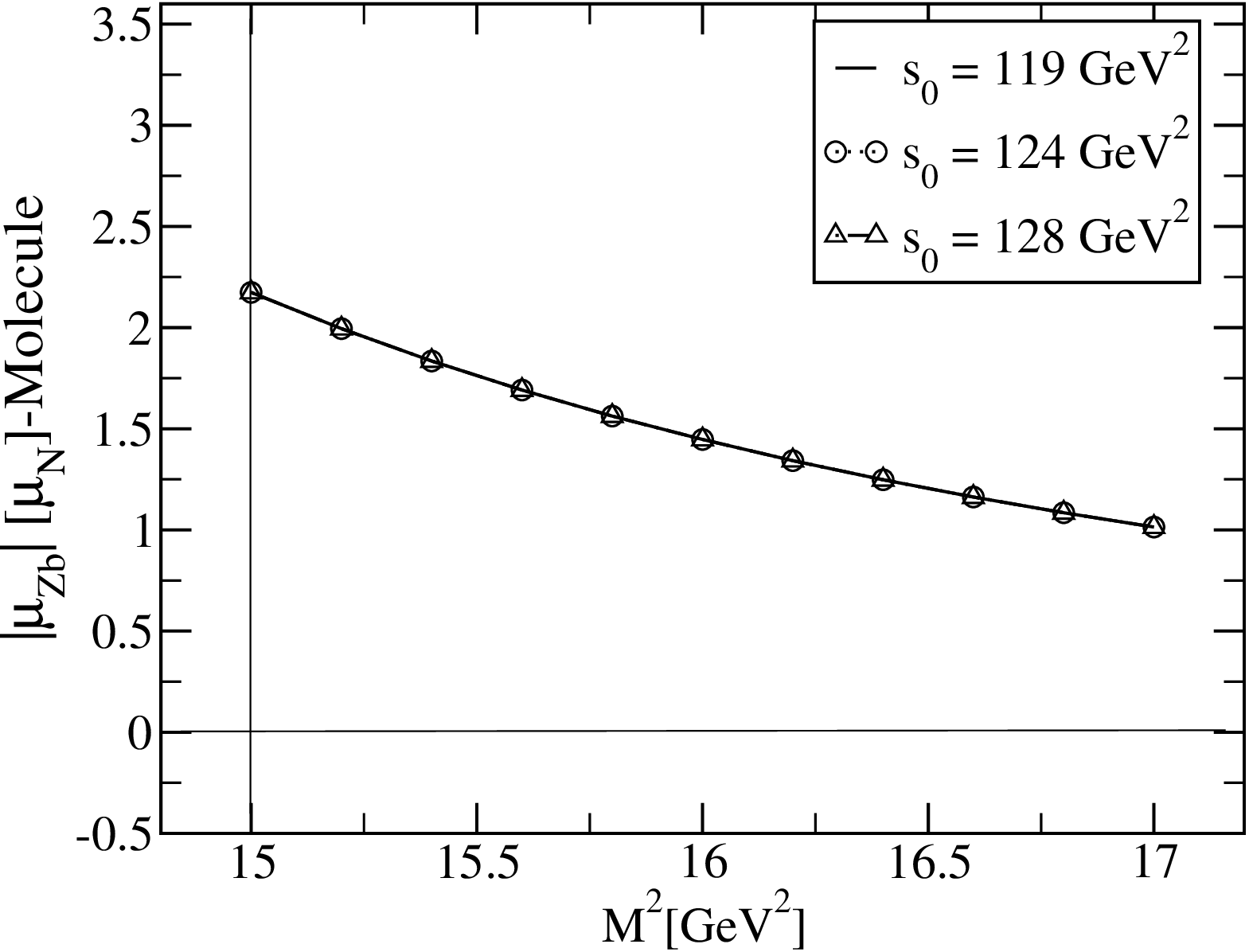}}
 \caption{ The dependence of the magnetic moment for $Z_b$ state; on the Borel parameter squared $M^{2}$
 at different fixed values of the continuum threshold.}
  \end{figure}

\begin{figure}
\centering
\subfloat[]{\label{fig:TCMM.eps}\includegraphics[width=0.4\textwidth]{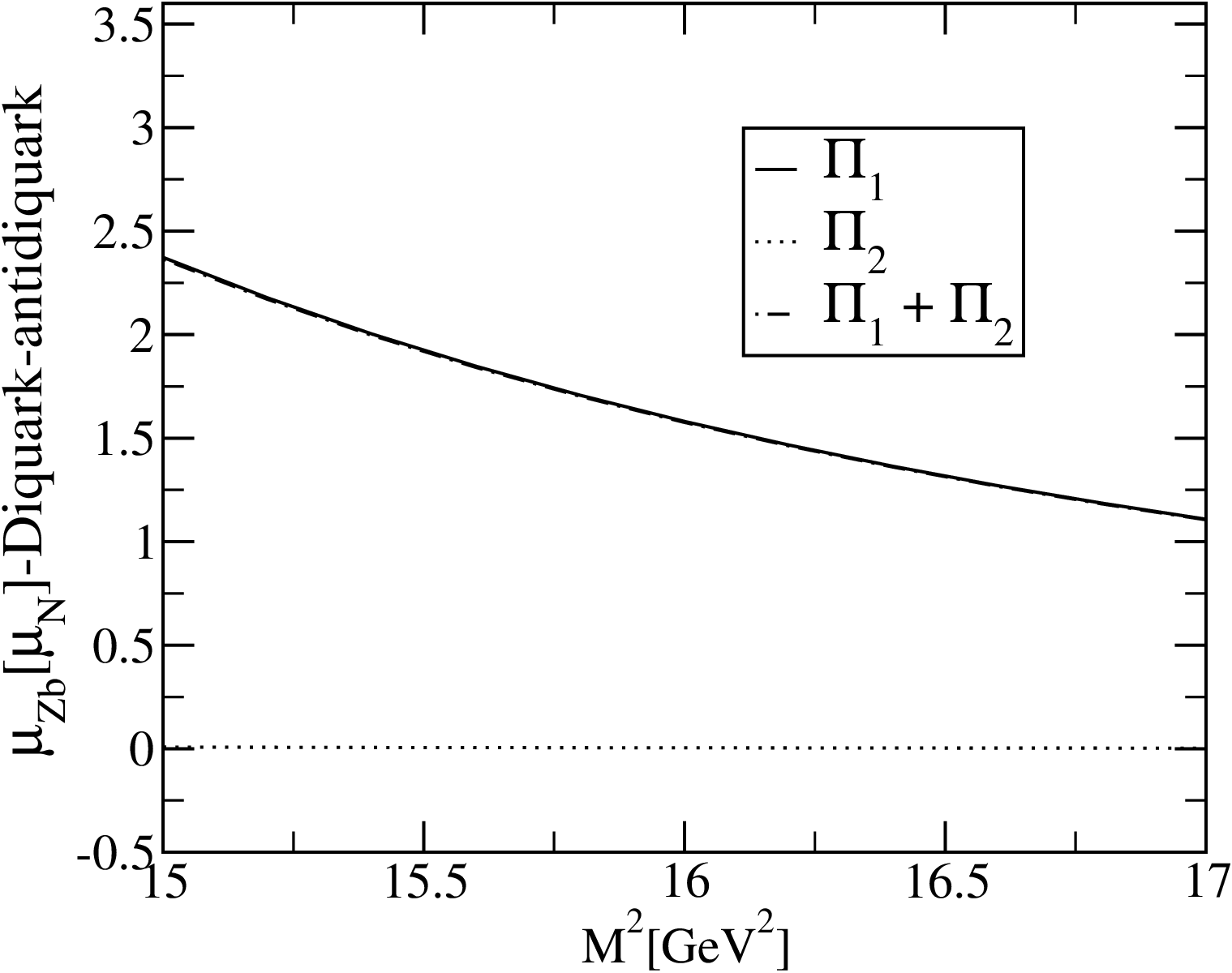}}\\
\vspace{0.5cm}
\subfloat[]{\label{fig:MCMM.eps}\includegraphics[width=0.4\textwidth]{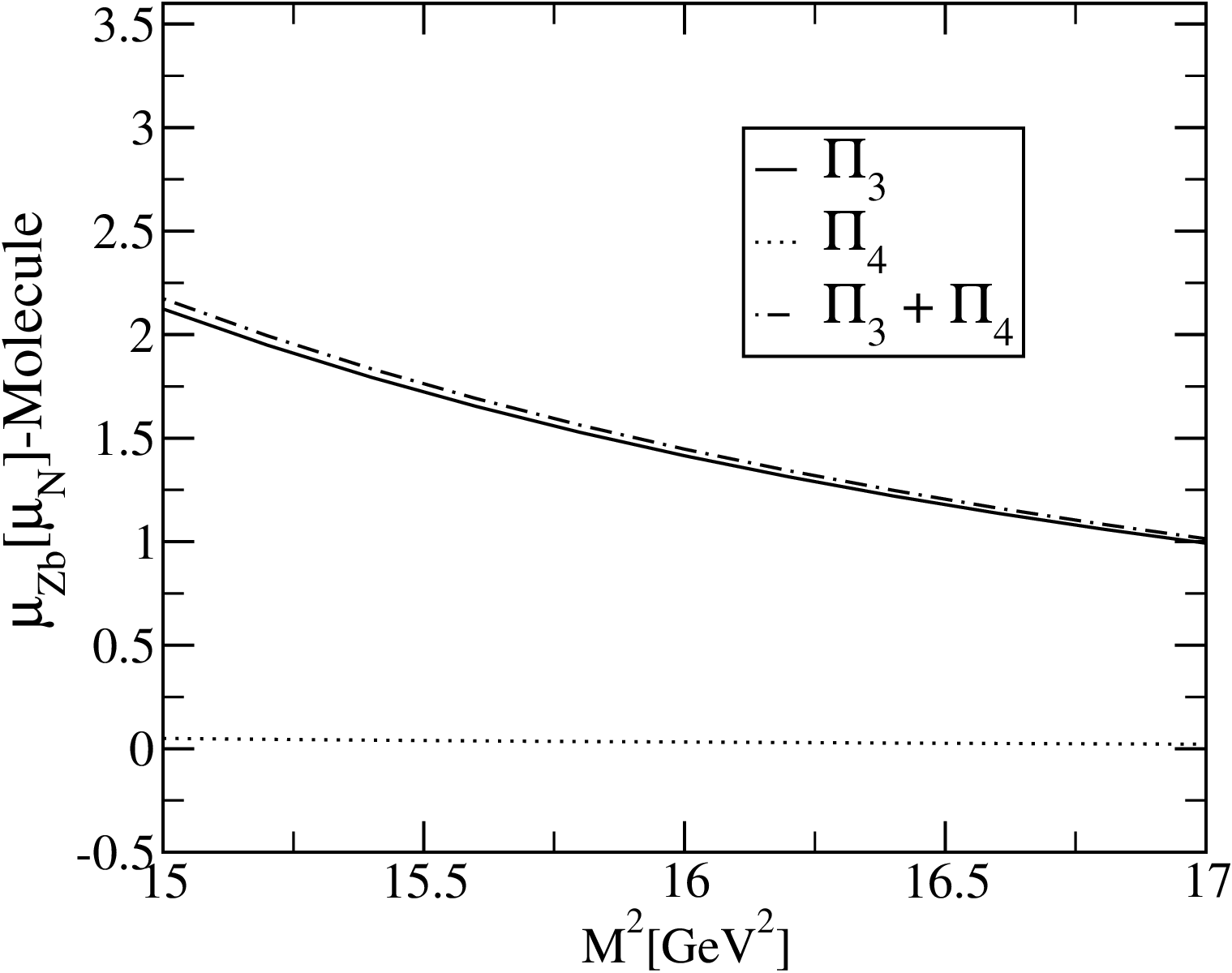}}
 \caption{ Comparison of the contributions to the magnetic moment with respect to $M^2$ at average
value of $s_0$.}
 \end{figure}

Our predictions on the numerical value of the magnetic dipole moment in both pictures are presented in Table I.
The errors in the results come from the variations in the  
calculations of the working regions of
 $M^2$ and from the uncertainties in the values of the input parameters as well as the photon DAs.
 We shall remark that the main source of uncertainties 
 is the variations with respect to variations of $M^2$.
 
\begin{table}[t]
	\addtolength{\tabcolsep}{10pt}
	\begin{center}
\begin{tabular}{ccccccc}\hline\hline
	   Picture &~~~~ $|\mu_{Z_b}|$&    \\\hline\hline
	Diquark-antidiquark&~~~~  1.73$\pm$ 0.63 & \\
	Molecule  &~~~~ 1.59$\pm$ 0.58  & \\\hline\hline
\end{tabular}
\end{center}
\caption{Results of the magnetic moment (in units of $\mu_N$) for $Z_b$ state.}
	\label{table}
\end{table}

In conclusion, we have computed the magnetic dipole moment
 of the $Z_b(10610)$ by modeling it as the
diquark-antidiquark and molecule states. In our calculations 
we have employed the light-cone QCD sum rule
in electromagnetic background field.
Although the central values of the  magnetic dipole moment obtained
via two pictures differ slightly from each other but they are consistent within the errors.
In Ref.~\cite{Agaev:2017lmc}, both the spectroscopic parameters 
and some  of the strong decays of the $ Z_b $ state have been studied using 
diquark-antidiquark interpolating current. Although the obtained mass in~\cite{Agaev:2017lmc} is in agreement with 
the experimental data, the result obtained for the width of $Z_b$ in the diquark-antidiquark picture 
in~\cite{Agaev:2017lmc} differ considerably from the experimental data.
They suggested, as a result, that the $Z_b$ state may not have a pure diquark-antidiquark structure.
When we combine the obtained results in the present study with those of the predictions on the mass obtained 
via both pictures in the literature and those result obtained  for the width of $Z_b$ in Ref.~\cite{Agaev:2017lmc}
 we conclude that both pictures can be considered for the internal structure of $Z_b$.
 May be a mixed current will be a better choice for interpolating this particle.
More theoretical and experimental studies are still needed to be performed in this respect.

Finally, the  magnetic dipole moment encodes important information about  the inner structure of particles 
and their geometric shape. 
The results obtained for the  magnetic dipole moment of $Z_b$ state in both the diquark-antidiquark and molecule pictures, within a factor 2, are of the same order of magnitude as the proton's magnetic moment and not such small that it appears hopeless to try to measure the value
 of the magnetic dipole moment of this state. By the recent progresses in the experimental side, we hope that we can measure the multipole moments of the newly founded exotic states, especially the $Z_b$ particle in future. 
Comparison of any experimental data on the  magnetic dipole moment of  $Z_b$ will be useful to gain
 exact knowledge on its quark organizations and will help us in the course of undestanding the structures 
 of the newly observed exotic states and their
 quantum chromodynamics.

\section{Acknowledgement}
This work has been supported by the Scientific and
Technological Research Council of Turkey (T\"{U}B\.{I}TAK)
under the Grant No. 115F183.

\appendix
\section*{Appendix A: Photon DAs and Wave Functions}
		        In this appendix, we present the definitions of the matrix elements of the
forms $\langle \gamma(q)\vel \bar{q}(x) \Gamma_i q(0) \ver 0\rangle$  
and $\langle \gamma(q)\vel \bar{q}(x) \Gamma_i G_{\mu\nu}q(0) \ver 0\rangle$ in terms of the photon
DAs and wave functions \cite{Ball:2002ps},

\begin{eqnarray*}
\label{esbs14}
&&\langle \gamma(q) \vert  \bar q(x) \gamma_\mu q(0) \vert 0 \rangle
= e_q f_{3 \gamma} \left(\varepsilon_\mu - q_\mu \frac{\varepsilon
x}{q x} \right) \int_0^1 du e^{i \bar u q x} \psi^v(u)
\nonumber \\
&&\langle \gamma(q) \vert \bar q(x) \gamma_\mu \gamma_5 q(0) \vert 0
\rangle  = - \frac{1}{4} e_q f_{3 \gamma} \epsilon_{\mu \nu \alpha
\beta } \varepsilon^\nu q^\alpha x^\beta \int_0^1 du e^{i \bar u q
x} \psi^a(u)
\nonumber \\
&&\langle \gamma(q) \vert  \bar q(x) \sigma_{\mu \nu} q(0) \vert  0
\rangle  = -i e_q \langle \bar q q \rangle (\varepsilon_\mu q_\nu - \varepsilon_\nu
q_\mu) \int_0^1 du e^{i \bar u qx} \left(\chi \varphi_\gamma(u) +
\frac{x^2}{16} \mathbb{A}  (u) \right) \nonumber \\ 
&&-\frac{i}{2(qx)}  e_q \bar qq \left[x_\nu \left(\varepsilon_\mu - q_\mu
\frac{\varepsilon x}{qx}\right) - x_\mu \left(\varepsilon_\nu -
q_\nu \frac{\varepsilon x}{q x}\right) \right] \int_0^1 du e^{i \bar
u q x} h_\gamma(u)
\nonumber \\
&&\langle \gamma(q) | \bar q(x) g_s G_{\mu \nu} (v x) q(0) \vert 0
\rangle = -i e_q \langle \bar q q \rangle \left(\varepsilon_\mu q_\nu - \varepsilon_\nu
q_\mu \right) \int {\cal D}\alpha_i e^{i (\alpha_{\bar q} + v
\alpha_g) q x} {\cal S}(\alpha_i)
\nonumber \\
&&\langle \gamma(q) | \bar q(x) g_s \tilde G_{\mu \nu}(v
x) i \gamma_5  q(0) \vert 0 \rangle = -i e_q \langle \bar q q \rangle \left(\varepsilon_\mu q_\nu -
\varepsilon_\nu q_\mu \right) \int {\cal D}\alpha_i e^{i
(\alpha_{\bar q} + v \alpha_g) q x} \tilde {\cal S}(\alpha_i)
\nonumber \\
&&\langle \gamma(q) \vert \bar q(x) g_s \tilde G_{\mu \nu}(v x)
\gamma_\alpha \gamma_5 q(0) \vert 0 \rangle = e_q f_{3 \gamma}
q_\alpha (\varepsilon_\mu q_\nu - \varepsilon_\nu q_\mu) \int {\cal
D}\alpha_i e^{i (\alpha_{\bar q} + v \alpha_g) q x} {\cal
A}(\alpha_i)
\nonumber \\
&&\langle \gamma(q) \vert \bar q(x) g_s G_{\mu \nu}(v x) i
\gamma_\alpha q(0) \vert 0 \rangle = e_q f_{3 \gamma} q_\alpha
(\varepsilon_\mu q_\nu - \varepsilon_\nu q_\mu) \int {\cal
D}\alpha_i e^{i (\alpha_{\bar q} + v \alpha_g) q x} {\cal
V}(\alpha_i) \nonumber\\
&& \langle \gamma(q) \vert \bar q(x)
\sigma_{\alpha \beta} g_s G_{\mu \nu}(v x) q(0) \vert 0 \rangle  =
e_q \langle \bar q q \rangle \left\{
        \left[\left(\varepsilon_\mu - q_\mu \frac{\varepsilon x}{q x}\right)\left(g_{\alpha \nu} -
        \frac{1}{qx} (q_\alpha x_\nu + q_\nu x_\alpha)\right) \right. \right. q_\beta
\nonumber \\ && -
         \left(\varepsilon_\mu - q_\mu \frac{\varepsilon x}{q x}\right)\left(g_{\beta \nu} -
        \frac{1}{qx} (q_\beta x_\nu + q_\nu x_\beta)\right) q_\alpha
-
         \left(\varepsilon_\nu - q_\nu \frac{\varepsilon x}{q x}\right)\left(g_{\alpha \mu} -
        \frac{1}{qx} (q_\alpha x_\mu + q_\mu x_\alpha)\right) q_\beta
\nonumber \\ &&+
         \left. \left(\varepsilon_\nu - q_\nu \frac{\varepsilon x}{q.x}\right)\left( g_{\beta \mu} -
        \frac{1}{qx} (q_\beta x_\mu + q_\mu x_\beta)\right) q_\alpha \right]
   \int {\cal D}\alpha_i e^{i (\alpha_{\bar q} + v \alpha_g) qx} {\cal T}_1(\alpha_i)
\nonumber \\ &&+
        \left[\left(\varepsilon_\alpha - q_\alpha \frac{\varepsilon x}{qx}\right)
        \left(g_{\mu \beta} - \frac{1}{qx}(q_\mu x_\beta + q_\beta x_\mu)\right) \right. q_\nu
\nonumber \\ &&-
         \left(\varepsilon_\alpha - q_\alpha \frac{\varepsilon x}{qx}\right)
        \left(g_{\nu \beta} - \frac{1}{qx}(q_\nu x_\beta + q_\beta x_\nu)\right)  q_\mu
\nonumber \\ && -
         \left(\varepsilon_\beta - q_\beta \frac{\varepsilon x}{qx}\right)
        \left(g_{\mu \alpha} - \frac{1}{qx}(q_\mu x_\alpha + q_\alpha x_\mu)\right) q_\nu
\nonumber \\ &&+
         \left. \left(\varepsilon_\beta - q_\beta \frac{\varepsilon x}{qx}\right)
        \left(g_{\nu \alpha} - \frac{1}{qx}(q_\nu x_\alpha + q_\alpha x_\nu) \right) q_\mu
        \right]      
    \int {\cal D} \alpha_i e^{i (\alpha_{\bar q} + v \alpha_g) qx} {\cal T}_2(\alpha_i)
\nonumber \\
&&+\frac{1}{qx} (q_\mu x_\nu - q_\nu x_\mu)
        (\varepsilon_\alpha q_\beta - \varepsilon_\beta q_\alpha)
    \int {\cal D} \alpha_i e^{i (\alpha_{\bar q} + v \alpha_g) qx} {\cal T}_3(\alpha_i)
\nonumber \\ &&+
        \left. \frac{1}{qx} (q_\alpha x_\beta - q_\beta x_\alpha)
        (\varepsilon_\mu q_\nu - \varepsilon_\nu q_\mu)
    \int {\cal D} \alpha_i e^{i (\alpha_{\bar q} + v \alpha_g) qx} {\cal T}_4(\alpha_i)
                        \right\}~,
\end{eqnarray*}
where $\varphi_\gamma(u)$ is the leading twist-2, $\psi^v(u)$,
$\psi^a(u)$, ${\cal A}(\alpha_i)$ and ${\cal V}(\alpha_i)$, are the twist-3, and
$h_\gamma(u)$, $\mathbb{A}(u)$, ${\cal S}(\alpha_i)$, ${\cal{\tilde S}}(\alpha_i)$, ${\cal T}_1(\alpha_i)$, ${\cal T}_2(\alpha_i)$, ${\cal T}_3(\alpha_i)$ 
and ${\cal T}_4(\alpha_i)$ are the
twist-4 photon DAs.
The measure ${\cal D} \alpha_i$ is defined as
\begin{eqnarray*}
\label{nolabel05}
\int {\cal D} \alpha_i = \int_0^1 d \alpha_{\bar q} \int_0^1 d
\alpha_q \int_0^1 d \alpha_g \delta(1-\alpha_{\bar
q}-\alpha_q-\alpha_g)~.\nonumber
\end{eqnarray*}

The expressions of the DAs entering into the above matrix elements are
defined as:

\begin{eqnarray}
\varphi_\gamma(u) &=& 6 u \bar u \left( 1 + \varphi_2(\mu)
C_2^{\frac{3}{2}}(u - \bar u) \right),
\nonumber \\
\psi^v(u) &=& 3 \left(3 (2 u - 1)^2 -1 \right)+\frac{3}{64} \left(15
w^V_\gamma - 5 w^A_\gamma\right)
                        \left(3 - 30 (2 u - 1)^2 + 35 (2 u -1)^4
                        \right),
\nonumber \\
\psi^a(u) &=& \left(1- (2 u -1)^2\right)\left(5 (2 u -1)^2 -1\right)
\frac{5}{2}
    \left(1 + \frac{9}{16} w^V_\gamma - \frac{3}{16} w^A_\gamma
    \right),
\nonumber \\
h_\gamma(u) &=& - 10 \left(1 + 2 \kappa^+\right) C_2^{\frac{1}{2}}(u
- \bar u),
\nonumber \\
\mathbb{A}(u) &=& 40 u^2 \bar u^2 \left(3 \kappa - \kappa^+
+1\right)  +
        8 (\zeta_2^+ - 3 \zeta_2) \left[u \bar u (2 + 13 u \bar u) \right.
\nonumber \\ && + \left.
                2 u^3 (10 -15 u + 6 u^2) \ln(u) + 2 \bar u^3 (10 - 15 \bar u + 6 \bar u^2)
        \ln(\bar u) \right],
\nonumber \\
{\cal A}(\alpha_i) &=& 360 \alpha_q \alpha_{\bar q} \alpha_g^2
        \left(1 + w^A_\gamma \frac{1}{2} (7 \alpha_g - 3)\right),
\nonumber \\
{\cal V}(\alpha_i) &=& 540 w^V_\gamma (\alpha_q - \alpha_{\bar q})
\alpha_q \alpha_{\bar q}
                \alpha_g^2,
\nonumber \\
{\cal T}_1(\alpha_i) &=& -120 (3 \zeta_2 + \zeta_2^+)(\alpha_{\bar
q} - \alpha_q)
        \alpha_{\bar q} \alpha_q \alpha_g,
\nonumber \\
{\cal T}_2(\alpha_i) &=& 30 \alpha_g^2 (\alpha_{\bar q} - \alpha_q)
    \left((\kappa - \kappa^+) + (\zeta_1 - \zeta_1^+)(1 - 2\alpha_g) +
    \zeta_2 (3 - 4 \alpha_g)\right),
\nonumber \\
{\cal T}_3(\alpha_i) &=& - 120 (3 \zeta_2 - \zeta_2^+)(\alpha_{\bar
q} -\alpha_q)
        \alpha_{\bar q} \alpha_q \alpha_g,
\nonumber \\
{\cal T}_4(\alpha_i) &=& 30 \alpha_g^2 (\alpha_{\bar q} - \alpha_q)
    \left((\kappa + \kappa^+) + (\zeta_1 + \zeta_1^+)(1 - 2\alpha_g) +
    \zeta_2 (3 - 4 \alpha_g)\right),\nonumber \\
{\cal S}(\alpha_i) &=& 30\alpha_g^2\{(\kappa +
\kappa^+)(1-\alpha_g)+(\zeta_1 + \zeta_1^+)(1 - \alpha_g)(1 -
2\alpha_g)\nonumber +\zeta_2[3 (\alpha_{\bar q} - \alpha_q)^2-\alpha_g(1 - \alpha_g)]\},\nonumber \\
\tilde {\cal S}(\alpha_i) &=&-30\alpha_g^2\{(\kappa -\kappa^+)(1-\alpha_g)+(\zeta_1 - \zeta_1^+)(1 - \alpha_g)(1 -
2\alpha_g)\nonumber +\zeta_2 [3 (\alpha_{\bar q} -\alpha_q)^2-\alpha_g(1 - \alpha_g)]\}.
\end{eqnarray}

Numerical values of parameters used in DAs are: $\varphi_2(1~GeV) = 0$, 
$w^V_\gamma = 3.8 \pm 1.8$, $w^A_\gamma = -2.1 \pm 1.0$, 
$\kappa = 0.2$, $\kappa^+ = 0$, $\zeta_1 = 0.4$, $\zeta_2 = 0.3$.

\end{document}